\def\nk{n_{\rm b}}

\def\snf{\sin f}
\def\csf{\cos f}
\def\cu{\cos u}
\def\su{\sin u}

\def\Pb{P_{\rm b}}

\def\rfr#1{Equation~(\ref{#1})}
\def\rfrs#1#2{Equations~(\ref{#1})~to~(\ref{#2})}

\def\derp#1#2{\rp{\partial{#1}}{\partial{#2}}}
\def\dert#1#2{\frac{{{\textrm{d}}}{#1}}{{{\textrm{d}}}{#2}}}

\def\virg#1{``#1"}

\def\eqi{\begin{equation}}
\def\eqf{\end{equation}}
\def\eqia{\begin{eqnarray}}
\def\eqfa{\end{eqnarray}}

\def\rp#1#2{{#1\over#2}}
\def\lb#1{\label{#1}}

\def\bds#1{\boldsymbol{#1}}

\def\cI{\cos I}
\def\sI{\sin I}

\def\ton#1{\left(#1\right)}
\def\qua#1{\left[#1\right]}
\def\grf#1{\left\{#1\right\}}

\documentclass[modern]{aastex}

\usepackage{hyperref}
\usepackage{float}
\usepackage{amsmath,textgreek,w-greek,wasysym}
\usepackage{amscd,lineno}
\usepackage{amssymb,dsfont}
\usepackage{graphicx,epsfig}
\usepackage{txfonts}
\usepackage{xr-hyper}

\RequirePackage{color}

\newcommand{\emaila}{lorenzo.iorio@libero.it}

\allowdisplaybreaks[1]

\begin{document}

\title{Constraining some $r^{-n}$ extra-potentials in modified gravity models with LAGEOS-type laser-ranged geodetic satellites}

\shortauthors{L. Iorio, M.L. Ruggiero}

\author{Lorenzo Iorio}
\affil{Ministero dell'Istruzione, dell'Universit\`{a} e della Ricerca (M.I.U.R.)  \\
Permanent address for correspondence: Viale Unit\`{a} di Italia 68, 70125, Bari (BA)}
\email{\emaila}

\author{Matteo Luca Ruggiero}
\affil{Politecnico di Torino \\
Corso Duca degli Abruzzi 24, 10129 - Torino. \\ INFN - Sezione di Torino \\
Via Pietro Giuria 1, 10125 - Torino}
\email{matteo.ruggiero@polito.it}

%

\begin{abstract}
We focus on several models of modified gravity which share the characteristic of leading to perturbations of the Newtonian potential  $\propto K_2~r^{-2}$ and
$\propto K_3~r^{-3}$. In particular, by using existing long data records of the LAGEOS satellites, tracked on an almost continuous basis with the Satellite Laser Ranging (SLR) technique, we set preliminary constraints on the free parameters $K_2,~K_3$  in a model-independent, phenomenological  way. We obtain $\left|K_2\right|\lesssim 2.1\times 10^6~\textrm{m}^4~\textrm{s}^{-2},~ -2.5\times 10^{12}~\textrm{m}^5~\textrm{s}^{-2}\lesssim K_3 \lesssim 4.1\times 10^{12}~\textrm{m}^5~\textrm{s}^{-2}.$
They are several orders of magnitude tighter than corresponding bounds existing in the literature inferred with different techniques and in other astronomical and astrophysical scenarios. Then, we specialize them to the different parameters characterizing the various models considered. The availability of SLR data records of increasing length and accuracy will allow to further refine and strengthen the present results.
\end{abstract}

\section{Introduction}\lb{intro}
Gravitational interactions are described with great accuracy within the framework of General Relativity (GR);  as a matter of fact, the predictions of the Einstein's theory of gravitation were verified with great accuracy during last century \citep{2014arXiv1409.7871W,2016Univ....2...23D} by means of experimental tests and observations that, in their great majority, were performed in the Solar System, where gravity can be adequately described by the weak-field and slow-motion approximation. There are, however, some noteworthy exceptions providing tests of GR in the strong gravity regime, such as those involving  binary pulsars	\citep{1979Natur.277..437T, pulsar2006}. Eventually, it is impossible not to mention the recent direct detection  of gravitational waves \citep{GW2016,GW2017,2016Univ....2...22C}, that are produced in the strong field regime and detected, on the Earth, as very small ripples in  spacetime.

There are  many evidences  on the current accelerated expansion of our Universe, coming from various observations, such as the type Ia supernovae,  the baryon acoustic oscillation, the cosmic microwave background \citep{Perlmutter:1997zf,Riess:1998cb,Tonry:2003zg, Knop:2003iy, Barris:2003dq, Riess:2004nr, Astier:2005qq,Eisenstein:2005su,Spergel:2006hy,Hinshaw:2012aka}. In the frame of the Standard Cosmological Model,  the best picture coming from these observations suggests that  the Universe content is  76\% \textit{dark energy}, 20\% \textit{dark  matter}, 4\% ordinary baryonic matter:  in order to match these observations with GR, we are forced to introduced \textit{dark entities} such as  matter and energy with peculiar characteristics. In particular, the \textit{dark energy}  is an exotic cosmic  fluid, which has not yet been detected directly, and which does not cluster as ordinary
matter; indeed, its behaviour closely resembles that of the cosmological constant  $\Lambda$, whose nature and origin are, however, difficult to explain \citep{Peebles:2002gy,Martin:2012bt}.  \textit{Dark matter} is supposed to be a cold and pressureless medium, whose distribution is that of a spherical halo around the galaxies. Actually, besides these difficulties in describing gravitational interactions at very large scales, there are problems with the foundations of General Relativity \citep{2016Univ....2...11V} which, as is, is not renormalizable and cannot be reconciled with a quantum description \citep{Stelle:1976gc,2016Univ....2...24L}: hence, gravitational interactions seem to stand apart from the Standard Model.

Taking into account these issues, there are  reasonable motivations  to consider extensions of GR (see the review paper by \citet{Berti:2015itd} for a description of various modified gravity models). One possible way to extend GR is to modify its geometric structure, generalizing Einstein's approach according to which \textit{gravity is geometry}: in doing so,  the richer geometric structure introduces the ingredients needed to match the observations. This is the case, for instance, of $f(R)$ gravity \citep{Capozziello:2011et, Sotiriou:2008rp, DeFelice:2010aj,2015Univ....1..123D}, Gauss-Bonnet \citep{ Nojiri:2005jg} or  $f(G)$ gravity \citep{DeFelice:2008wz}, scalar-tensor gravity \citep{ Naruko:2015zze,Saridakis:2016ahq}, massive gravity \citep{deRham:2014zqa}.
Interestingly enough, GR and $f(R)$ gravity are subclasses of the so-called Horndeski theory \citep{Horndeski:1974wa}, which is the most general scalar-tensor theory whose action has higher derivatives of the scalar field $\phi$, but leads to second order differential equations, thus avoiding the Ostrogradsky instability. A different strategy to the extension of GR can be fulfilled starting from its equivalent formulation in terms of Teleparallel Gravity (TEGR) \citep{pereira,Maluf:2013gaa,2016Univ....2...19M}, thus obtaining $f(T)$ gravity \citep{Ferraro:2008ey,Linder:2010py,Cai:2015emx}.

However, it is manifest that any model of modified gravity should be in agreement with the known tests of GR, in particular in the Solar System: every extended theory of gravity is expected to reproduce GR in a suitable weak-field limit. As a consequence, modified gravity models must have correct Newtonian and post-Newtonian limits and, up
to intermediate scales, the deviations from the GR predictions can be considered as
perturbations; in other words,  these theories should have spherically symmetric solutions with gravitational Newtonian potential $U_\textrm{N}=-{GM}/{r}$ to which they add specific model-dependent perturbations, whose parameters, on the other hand, can be constrained by Solar System tests. For instance, this has been done for scalar-tensor theories and, more in general, Horndeski theory   \citep{Clifton:2011jh,Bhattacharya:2016naa}, $f(R)$ \citep{Berry:2011pb,Capozziello:2007ms,Capozziello:2006jj,Capone:2009xk,Allemandi:2006bm,Ruggiero:2006qv}, $f(T)$ \citep{2012MNRAS.427.1555I,2016PhRvD..93j4034F,Lin:2016nvj}.  In \citet{Iorio:2016sqy}  the Schwarzschild-de Sitter  solution arising in various models of modified gravity has been constrained by Solar System data.

In this paper, we aim at setting preliminary constraints of some models of modified gravity by means of the Earth's geodetic satellites of LAGEOS family tracked on an almost continuous basis with the Satellite Laser Ranging (SLR) technique \citep{Combrinck2010} to a $\simeq~\textrm{cm}$ accuracy level. In particular, we are going to focus on those whose perturbations with respect to the Newtonian potential fall off as the square or the cube of the distance from the central mass $M$.

The paper is organized as follows. After briefly reviewing the origin of these models in Section~\ref{palle}, in Section~\ref{secU2} we deal with a $r^{-2}$ extra-potential, while Section~\ref{secU3} is devoted to the $r^{-3}$ case. In Section~\ref{fine}, we summarize our findings and offer our conclusions including the constraints on the models' parameters inferred with laser data from geodetic Earth's satellites. Basic notations and definitions used throughout the text are collected in Section~\ref{appen}. The analytical calculational approach adopted is detailed in Section~\ref{analy}. Section~\ref{tabfig} contains tables and figures.
\section{Spherically symmetric solutions for modified gravity models}\lb{palle}
In this Section, we are going to review the weak-field solutions that, in some models of modified gravity, can be used to describe the dynamics in the Solar System. In doing so, we assume that the generic time-time component of the spacetime metric is in the form
\eqi
g_{00} \simeq  1 + h_{00}, \label{eq:g00}
\eqf
where $h_{00}$ is a {small perturbation} of the Minkowski spacetime. The gravitational potential
\eqi
U = \rp{c^2 h_{00}}{2} \label{eq:h00}
\eqf
consists of the sum of two contributions
\eqi
U = U_\textrm{N} + \Delta U_n,~n=2,3, \label{eq:defU}
\eqf
i.e. the Newtonian potential $\displaystyle U_\textrm{N}=-\frac{GM}{r}$  and the additional term $\Delta U_n$, which is an extra-potential peculiar to the modified gravity model considered. We assume that $|\Delta U_{n}|\ll |U_{N}|$, so that it can be treated as a perturbation. Furthermore, we use the following notation
\eqi
\Delta U_2=\rp{K_2}{r^2},~\qua{K_2}=\textrm{L}^4~\textrm{T}^{-2}\label{eq:defU2}
\eqf
for extra-potentials falling off as $\displaystyle \sim \frac{1}{r^{2}}$ and
\eqi
\Delta U_3=\rp{K_3}{r^3},~\qua{K_3}=\textrm{L}^5~\textrm{T}^{-2} \label{eq:defU3}
\eqf
for those falling off as $\displaystyle \sim \frac{1}{r^{3}}$.
\subsection{The $r^{-2}$ extra-potentials}
Here, we focus on some models of modified gravity leading to an additional term proportional to $r^{-2}$. To begin with, we remember that, in classical GR, the Reissner-Nordstr\"om metric \citep{wald2010general}, which describes the gravitational field of charged, non-rotating spherically symmetric body, has just an $r^{-2}$ term related to the charge $Q$ of the source. In this case, we may write
\eqi
\Delta U_{2} = \frac{GQ^{2}}{8\uppi\varepsilon_0 c^{2}r^{2}},
\eqf  and
\eqi
K_{2}=\frac{GQ^{2}}{8\uppi\varepsilon_0 c^{2}}\lb{ReNo}.
\eqf Constraints on the net electric charge $Q$ of  astronomical and astrophysical objects  have been set by \citet{Iorio:2011aa}; in particular, the constraint for the Earth charge is $|Q| \lesssim 4 \times 10^{13}$ C, obtained by studying the GRACE mission  \citep{tapley2004gravity} around the Earth.

\citet{2015PhRvD..91j4014R}, in the framework of $f(T)$ gravity, studied  weak-field spherically symmetric solutions for Lagrangians in the form   $f(T)=T+\alpha T^n$, where $\alpha$ is a small constant, whose dimensions are $\qua{\alpha}=\textrm{L}^2$, parametrizing the departure of these theories from $GR$, and $|n| \neq 1$. Among their results, the case with $n=2$, corresponding to the Lagrangian $f(T)=T+\alpha T^{2}$,  is interesting since every general Lagrangian reduces to this form, in first approximation. Cosmological constraints on these models of modified gravity have ben set by \citet{Nunes:2016qyp,2018XU} showing that they are consistent with the observations. The corresponding extra-potential turns out to be $\displaystyle \Delta U_{2}=-\frac{16\alpha c^{2}}{r^{2}}$, and $\displaystyle K_{2}= -16 \alpha c^{2}$.  A quadratic Lagrangian in $f(T)$ gravity was considered also by \citet{2012MNRAS.427.1555I}, using a different approach, to  obtain a weak-field spherically symmetric solution for the gravitational field in the Solar System. This lead to a slightly different parameterization: $K_2=-3\alpha c^2$.

In  Einstein-Gauss-Bonnet gravity, the Maeda-Dadhich solution \citep{Maeda:2006hj, Bhattacharya:2016naa} has an extra potential in the form $\displaystyle \Delta U_{2}=\frac{2G^{2}M^{2}\tilde q}{c^{2}r^{2}}$, where $\tilde q$ is a dimensionless parameter, whose best constraint $|\tilde q| \lesssim 0.024$  has been obtained by perihelion precession \citep{Bhattacharya:2016naa}; in this case $\displaystyle K_{2}=\frac{2G^{2}M^{2}\tilde q}{c^{2}}$.

\citet{Ali:2015tva} obtained a quantum corrected Schwarzschild metric, starting  from a quantum Raychaudhuri equation (QRE) (see also \citet{Jusufi:2016sym}); in this context, the extra potential is $\displaystyle \Delta U_{2}=\frac{\hbar G \eta}{2c r^{2}}$, where $\eta$ is a dimensionless constant; in this case $\displaystyle K_{2}= \frac{\hbar G \eta}{2c }$.

\subsection{The $r^{-3}$ extra-potentials}
Here, we focus on some modified gravity models whose extra-potential is proportional to $r^{-3}$.

\citet{Bonanno:2000ep}, using the renormalization group approach, obtained a modification of the Schwarzschild metric whose asymptotical behaviour contains a perturbation $\displaystyle \Delta U_{3}=\frac{G^{2}M \omega}{c^{3}r^{3}}$. In this model the parameter $\displaystyle \omega=\frac{167 \hbar}{30\pi}$ is a constant which  encodes the quantum effects \citep{Jusufi:2016sym}; actually, there are no free parameters in this model, so it cannot be constrained by observations.

The Sotiriou-Zhou solution \citep{Sotiriou:2014pfa} is obtained starting from the coupling of a scalar field $\phi$ with the Gauss-Bonnet invariant; however, it is important to emphasize \citep{Bhattacharya:2016naa} that, in this case, such a  solution is valid for black hole or may describe wormholes \citep{Kanti:2011jz,Kanti:2011yv}, so it not suitable for properly describing the spacetime around, say, a star like the Sun. Nonetheless, because of its interest in describing the dynamics around, e.g., the galactic black hole, we mention it here. Now, the perturbation is $\displaystyle \Delta U_{3}=\frac{GMP^{2}}{12r^{3}}$, where $P$ is a constant whose dimensions are $\qua P=\textrm{L}^{2}$. It represents the charge associated to the scalar field.
Indeed, in what follows we will not constrain this model, since our analysis is based on the  motion of the Earth's geodetic satellites. It is interesting to point out that, as shown by \citet{Antoniou:2017hxj,Antoniou:2017acq}, the Sotiriou-Zhou solution is a special case of linear coupling between the scalar field with the Gauss-Bonnet invariant; the more general case is considered in the aforementioned papers.

In the framework of string theory, there are closed string excitations leading to a second rank antisymmetric tensor field, known as the Kalb-Ramond field which, from a certain viewpoint, generalizes the electromagnetic potential \citep{Chakraborty:2016lxo}.  It has been suggested    that this field may have an impact on the four dimensional spacetime: in particular (see \citet{Chakraborty:2016lxo} and references therein) if the
Kalb-Ramond field is present in four spacetime dimension, there is the extra-potential $\displaystyle \Delta U_{3}=-\frac{GMb}{3r^{3}}$. Here, $b$ is  the Kalb-Ramond parameter with dimensions $\qua b=\textrm{L}^{2}$; accordingly, we have $\displaystyle K_{3} = -\frac{GMb}{3}$.

For the sake of completeness, we mention here that similar extra-potentials proportional to $r^{-3}$ have been obtained also in different models of modified gravity, which are however effective at the particle physics scales   \citep{2001PhRvD..64g5010F,2003ARNPS..53...77A, 1999PhRvL..83.4690R,2007PhRvL..98m1104A,1987PhLA..125..405M,1998PhRvD..58i6006F,1999PhRvD..59g5009F,2006JHEP...11..005D,2009PrPNP..62..102A} and, hence, cannot be constrained  using our approach.

\section{The constraints on the $r^{-2}$ extra-potential}\lb{secU2}
From \rfr{eq:defU2}, the perturbing radial acceleration
\eqi
A_2 = -2 \rp{K_2}{r^3}\lb{A2}
\eqf
arises.

%

\begin{figure*}
\centerline{
\vbox{
\begin{tabular}{cc}
\epsfysize= 5.00 cm\epsfbox{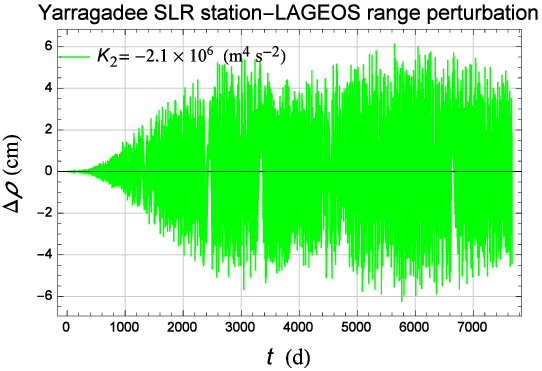}&
\epsfysize= 5.00 cm\epsfbox{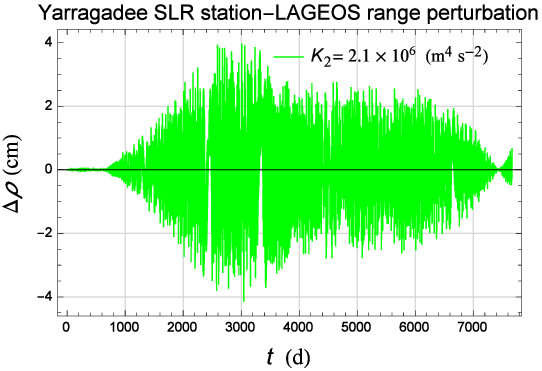}\\
\end{tabular}
}
}
\caption{Numerically produced time series of the perturbation $\Delta\rho(t)$ of the range $\rho$ between the Earth-based SLR station 7090 (Yarragadee, Australia) and the LAGEOS satellite due to \rfr{eq:defU2} for $K_2 = \mp 2.1\times 10^6~\textrm{m}^4~\textrm{s}^{-2}$ as the difference of two numerical integrations of the satellites's equations of motion in rectangular Cartesian coordinates with and without \rfr{A2} over the same time span 21 yr long (1993-2014) of Figure 2 of \citet{2016JGeod..90.1371A}. The station coordinates were retrieved from https://ilrs.cddis.eosdis.nasa.gov/network/stations/active/YARL$\_$general.html, while the HORIZONS Web-interface by NASA JPL (https://ssd.jpl.nasa.gov/horizons.cgi) was used to retrieve the initial state vector of LAGEOS. }\label{figura1}
\end{figure*}

According to Figure~2 of \citet{2016JGeod..90.1371A}, the range residuals $\delta\rho(t)$ of LAGEOS obtained by fitting a complete set of dynamical and measurement models of several standard gravitational and non-gravitational effects to precise ranging measurements collected from 1993 to 2014 by some Earth-based SLR stations are at the $\simeq 2-5~\textrm{cm}$ level. {More precisely, the directly observable quantities with the SLR technique are the measurements of the two-way time-of-flight of the electromagnetic radiation bounced back by the retroreflectors which entirely cover the LAGEOS surface. They are then corrected for additional delays due to the atmosphere, satellite centre-of-mass, the Shapiro delay, etc. As an outcome, a time series of station-satellite range measurements $\rho_\textrm{O}(t)$ performed at various epochs is obtained; it is dubbed with \virg{O}, which stands for \virg{Observable}. The next step consists of an accurate mathematical modeling of the entire range measurement process, including the satellite's dynamics, the propagation of the laser pulses and the instruments' functioning and measurement procedure; as a consequence, a time series  \virg{C} of  station-satellite ranges $\rho_\textrm{C}(t)$, calculated at the same epochs of the measured ones, is produced, usually with numerical techniques. At this stage, it should be kept in mind that  the models used in this step are, in general, inaccurate because of a number of reasons: the mathematical form of some of their parts can be partly or totally wrong, the physical parameters entering them are known with unavoidably limited accuracy, some more or less fundamental pieces of Nature, like, e.g., this or that dynamical accelerations affecting the satellite's motion, are not modeled at all. Then, the  time series $\rho_\textrm{C}(t)$ is fit to $\rho_\textrm{O}(t)$ in a least-square way by estimating a huge number of solve-for parameters $\grf{\textrm{p}}$. Usually they include, among others, also quantities in terms of which the gravitational environment is expressed like, e.g., the primary's mass, multipole moments, etc. Finally, $\rho_\textrm{C}(t)$ is re-calculated at the same epochs of the measurements of $\rho_\textrm{O}(t)$ by means of the previously estimated parameters $\grf{\textrm{p}}$. Thus, a post-fit time series $\rho_\textrm{C}^\textrm{pf}\ton{t; \grf{\textrm{p}}}$ is generated and subtracted from $\rho_\textrm{O}(t)$ in order to obtain the time series of the post-fit range residuals $\delta\rho(t)= \rho_\textrm{C}^\textrm{pf}\ton{t; \grf{\textrm{p}}} - \rho_\textrm{O}(t)$. If the whole data reduction went smooth and the models were adequate, the temporal pattern of $\delta\rho(t)$ should look like a rather uniform band, without any discernable peculiar feature like, say, a secular trend or a harmonic signature. The mean value of $\delta\rho(t)$ is smaller than its standard deviation or of any other statistical measure of its scatter which should not excess too much the size of the measurement errors; the ultimate goal of an accurate modeling is, indeed, to push the accuracy of the post-fit residuals down to the level of the measurement errors themselves. In principle, the post-fit residuals account, among other things, also for any unmodeled or mismodeled feature of motion, and can be used to put constraints on it by setting the largest admissible value compatible with the actual width of the range residuals.
}
By straightforwardly comparing our Figure~\ref{figura1}, which depicts a numerically produced time series of the range perturbation induced by \rfr{A2} on the distance from the Yarragadee station to LAGEOS, with Figure~2 of \citet{2016JGeod..90.1371A}, it is possible to preliminarily infer
\eqi
\left|K_2\right|\lesssim 2.1\times 10^6~\textrm{m}^4~\textrm{s}^{-2}\lb{K2}
\eqf
{in the sense that larger values of $\left|K_2\right|$ would generate a simulated  signature with an amplitude $\Delta\rho$ exceeding the $\simeq 2-5~\textrm{cm}$ level of Figure~2 in \citet{2016JGeod..90.1371A}. In other words, if $\left|K_2\right|$ were larger than \rfr{K2}, the theoretical time series of its range perturbation would not stay within the margins of the experimental post-fit residuals of Figure~2 in \citet{2016JGeod..90.1371A} which, in principle, fully account also for it since no unconventional dynamics was modeled at all.}
In the parameterization of \citet{2012MNRAS.427.1555I,2013MNRAS.433.3584X}
\eqi
K_2\rightarrow -3c^2\alpha,
\eqf
the bound of \rfr{K2} corresponds to
\eqi
\left|\alpha\right|\lesssim 7.79 \times 10^{-12}~\textrm{m}^2,\lb{bound1}
\eqf
while, from \citep{2015PhRvD..91j4014R,2015JCAP...08..021I}
\eqi
K_2\rightarrow -16 c^2 \alpha,
\eqf
one gets
\eqi
\left|\alpha\right|\lesssim 1.46 \times 10^{-12}~\textrm{m}^2.\lb{bound2}
\eqf
It must be noted that \rfr{bound1} is about $16-14$ orders of magnitude better than the bounds previously obtained in \citet{2012MNRAS.427.1555I,2013MNRAS.433.3584X}, while \rfr{bound2} improve the results in \citet{2015PhRvD..91j4014R,2015JCAP...08..021I} by about $14-11$ orders of magnitude.
It is interesting to note that \rfr{bound2} is even smaller than the lower bound
\eqi
\left|\alpha\right|_\textrm{min} = 8.07\times 10^{-9}~\textrm{m}^2
\eqf
reported in \citet{2016PhRvD..93j4034F} by about 4 orders of magnitude.
In order to obtain their tightest constraints, both \citet{2012MNRAS.427.1555I,2013MNRAS.433.3584X} and \citet{2015PhRvD..91j4014R,2015JCAP...08..021I,2016PhRvD..93j4034F} used as observables the most recent observational constraints available at that times on the secular perihelion precessions of some inner planets in the field of the Sun by comparing them with the theoretical predictions for the anomalous pericenter precessions due to \rfr{A2}.

As for the other models of modified gravity,  for the charge in the Reissner-Nordstr\"om metric we obtain $|Q| \lesssim 7.93 \times 10^{11}$ C which is about two orders of magnitude better than the bounds obtained in by \citet{Iorio:2011aa}. Our bound on the Maeda-Dadhich solution parameter is $|\tilde q| \lesssim 5.94 \times 10^{-7}$, which is about six orders of magnitude better than the previous best constraint obtained by \citet{Bhattacharya:2016naa}. Eventually, for the
\citet{Ali:2015tva} parameter, we obtain $|\eta| \lesssim 1.79 \times 10^{59}$: we remember that this model is determined by quantum corrections, hence the scale where these corrections are supposed to be effective is quite different than the one we are testing here.

We remark that \citet{2016JGeod..90.1371A} did not explicitly model any modified model of gravity. Thus, \rfr{A2}, if really existent in Nature, may have been partially absorbed in the usual parametric estimation of the standard data reduction procedure and, at least to a certain extent, removed from the time series displayed in Figure~2 of \citet{2016JGeod..90.1371A}. As a consequence, the bound of \rfr{K2} may turn out to be somewhat optimistic, i.e. too tight. Anyway, it is not possible to a-priori quantify such a putative partial removal just on speculative grounds. Only a dedicated re-analysis of the same data set used in \citet{2016JGeod..90.1371A} by explicitly modeling \rfr{A2} and estimating $K_2$ along with the other usual solve-for parameters could, perhaps, effectively assess the impact of using straightforwardly our Figure~\ref{figura1} in a direct comparison with Figure~2 of \citet{2016JGeod..90.1371A}. On the other hand, it must also be noted that \rfr{K2} was conservatively inferred by assuming that range residuals by \citet{2016JGeod..90.1371A} were entirely due to \rfr{A2} itself. If, instead, they were to be partly attributed to other unmodelled/mismodelled conventional physical effects, the remaining putative contribution of \rfr{A2} to Figure~2 of \citet{2016JGeod..90.1371A} would yield a bound on $K_2$ even smaller than \rfr{K2} itself. Moreover, it is also possible that, even by explicitly modeling and solving for $K_2$ in a dedicated re-analysis of the SLR observations, the resulting constraints on it may still be affected by any other possible unmodeled/mismodeled acceleration, both of standard and exotic nature. Indeed, in standard practice, it is not possible to determine everything; a selection of the dynamical effects to be modeled and of their parameters which can be practically estimated is always unavoidably made in real data reductions. Thus, the effect of any sort of \virg{Russell teapots} may well still creep into the desired solved-for values of $K_2$ estimated in a full covariance analysis. Furthermore, it cannot be kept silent that the present approach has been-and is-largely adopted in the current literature (e.g. by \citet{2012MNRAS.427.1555I,2013MNRAS.433.3584X,2015PhRvD..91j4014R,2015JCAP...08..021I}) to infer bounds on any sort of non-standard modified models of gravity by using completely different kinds of data ranging from planetary observations to pulsar timing previously processed by other teams who modelled only standard physics inasmuch the same way as we did here. In any case, even if the bounds of \rfr{K2} and \rfrs{bound1}{bound2} were to be up to one order of magnitude weaker, nonetheless they would represent a quite remarkable improvement with respect to the planetary ones.

By applying the computational scheme outlined in Section~\ref{analy} to the perturbing radial acceleration of \rfr{A2}, it is possible to obtain the corresponding radial, transverse and normal orbital perturbations over an integer number $j$ of revolutions; they turn out to be
\begin{align}
\Delta R & = 0\lb{DR2}, \\ \nonumber \\
\Delta T & = -\rp{j 2\uppi K_2}{\mu \ton{1 + e\cos f_0}},~j\in\mathbb{N}^+,~j\geq 1\lb{DT2}\\ \nonumber \\
\Delta N & = 0.\lb{DN2}
\end{align}
The results of \rfrs{DR2}{DN2} are exact in $e$ since no a-priori simplifying approximations were adopted in deriving them. Furthermore, \rfrs{DR2}{DN2} can be used to infer other independent bounds on $K_2$ by comparing them with the time series of the residuals $\delta R(t),\delta T(t),\delta N(t)$ of the LAGEOS and LAGEOS II spacecraft covering $\Delta t = 13~\textrm{yr}$ produced by \citet{slr2006} and displayed in their Figure~2 and Figure~12.
The resulting preliminary constraints turn out to be about one-two orders of magnitude weaker than that of \rfr{K2}. Indeed, since in the case treated in \citet{slr2006}, it is
\eqi
j\simeq 30,731
\eqf
and the RMS of the transverse orbital components of the two LAGEOS satellites are of the order of a few cm, as per Table~5 and Table~7 of \citet{slr2006}, \rfr{DT2} returns
\eqi
\left|K_2\right|\lesssim \ton{3.4-9.1}\times 10^7~\textrm{m}^4~\textrm{s}^{-2}.\lb{K2LL2}
\eqf
Since also \citet{slr2006} modeled just standard physics, the same caveat previously described for \rfr{K2} holds to the bounds of \rfr{K2LL2} as well.
\section{The constraints on the $r^{-3}$ extra-potential}\lb{secU3}
From \rfr{eq:defU3}, the extra-acceleration
\eqi
A_3 = -3 \rp{K_3}{r^4}\lb{A3}
\eqf
arises.

\begin{figure*}
\centerline{
\vbox{
\begin{tabular}{cc}
\epsfysize= 5.00 cm\epsfbox{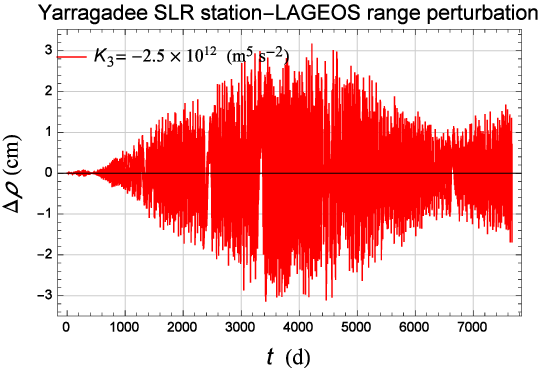}&
\epsfysize= 5.00 cm\epsfbox{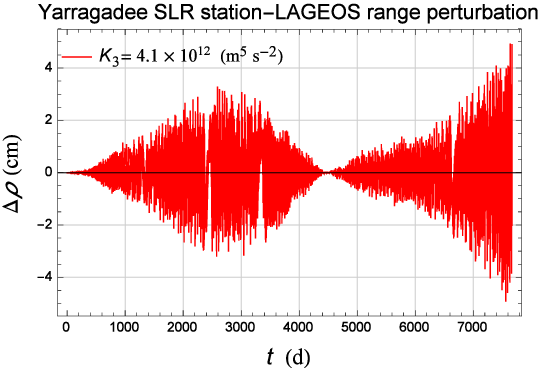}\\
\end{tabular}
}
}
\caption{Numerically produced time series of the perturbation $\Delta\rho(t)$ of the range $\rho$ between the Earth-based SLR station 7090 (Yarragadee, Australia) and the LAGEOS satellite due to \rfr{eq:defU3} for $K_3 = -2.5\times 10^{12}~\textrm{m}^5~\textrm{s}^{-2}~(\textrm{left~panel}),~K_3 =4.1\times 10^{12}~\textrm{m}^5~\textrm{s}^{-2}~(\textrm{right~panel})$ as the difference of two numerical integrations of the satellites's equations of motion in rectangular Cartesian coordinates with and without \rfr{A3} over the same time span 21 yr long (1993-2014) of Figure 2 of \citet{2016JGeod..90.1371A}. The station coordinates were retrieved from https://ilrs.cddis.eosdis.nasa.gov/network/stations/active/YARL$\_$general.html, while the HORIZONS Web-interface by NASA JPL (https://ssd.jpl.nasa.gov/horizons.cgi) was used to retrieve the initial state vector of LAGEOS. }\label{figura2}
\end{figure*}

By proceeding as in Section~\ref{secU2}, a straightforward comparison of the range residuals $\delta\rho(t)$ of LAGEOS produced by \citet{2016JGeod..90.1371A} with the numerically computed time series of the range perturbation $\Delta\rho(t)$ due to \rfr{A3}, displayed in Figure~\ref{figura2}, allows to preliminarily infer
\eqi
-2.5\times 10^{12}~\textrm{m}^5~\textrm{s}^{-2}\lesssim K_3 \lesssim 4.1\times 10^{12}~\textrm{m}^5~\textrm{s}^{-2}.\lb{boundK3}
\eqf

On using these bounds, we obtain the following constraints on the Kalb-Ramond parameter: $|b| \lesssim 0.038 $ m$^{2}$.

%
The bounds in \rfr{boundK3} are about four orders magnitude tighter than those released in \citet{2012AnP...524..371I} referring to the Earth's field.
The radial, transverse and normal orbital shifts after $j$ orbital revolutions are
\begin{align}
\Delta R & = 0, \\ \nonumber \\
\Delta T & = -\rp{j6\uppi K_3}{\mu a \ton{1 - e^2}\ton{1 + e\cos f_0}},~j\in\mathbb{N}^+,~j\geq 1\lb{DT3} \\ \nonumber \\
\Delta N & = 0.
\end{align}
By using \rfr{DT3} and the RMS of the transverse residuals $\delta T(t)$ of LAGEOS and LAGEOS II published in \citet{slr2006}, it can be obtained
\eqi
\left|K_3\right|\lesssim  \ton{1.4-3.7}\times 10^{14}~\textrm{m}^5~\textrm{s}^{-2};\lb{K3LL2}
\eqf
such figures are about two orders of magnitude weaker than the bounds of \rfr{boundK3}.
\section{Summary and conclusions}\lb{fine}
We exploited existing accurate time series of station-spacecraft range residuals of the geodetic satellites of the LAGEOS family to preliminary put constraints in the field of Earth on some modified models of gravity falling as $r^{-n},~n=2,3$. After having constrained their phenomenological parameters $K_2,~K_3$ without making any assumptions on the theoretical frameworks giving rise to them, we translated such bounds in terms of the parameters of some specific models yielding  $r^{-n},~n=2,3$ extra-potentials. Our results are summarized in  Table~\ref{tabletot}. Although necessarily preliminary because the modified models considered here are not explicitly modeled in all the currently available SLR data reductions, the resulting constraints turn out to be much tighter than other ones existing in the literature, especially in those cases in which $K_2,~K_3$ are independent of the source of the gravity field. Thus, they show the great potential of the approach proposed here. To this aim, it is important to stress that the lifetime of the LAGEOS satellites, which are tracked on an almost continuous basis since decades in view of their great importance in several geodetic studies, is of the order of $\approx 10^5$ yr. The availability of data records of ever increasing length should allow to further improve and make more robust the present constraints in a foreseeable future.
\appendix
\section{Notations and definitions}\lb{appen}
Here, some basic notations and definitions used in the text are presented
\begin{description}
\item[] $G:$ Newtonian constant of gravitation
\item[] $c:$ speed of light in vacuum
\item[] $M:$ mass of the primary
\item[] $\mu\doteq GM:$ gravitational parameter of the primary
\item[] $\bds{r}:$ position vector of the satellite
\item[] $r:$ distance of the satellite to the primary
\item[] $a:$  semimajor axis
\item[] $\nk \doteq \sqrt{\mu a^{-3}}:$   Keplerian mean motion
\item[] $\Pb = 2\uppi \nk^{-1}:$ Keplerian orbital period
\item[] $e:$  eccentricity
\item[] $p\doteq a(1-e^2):$  semilatus rectum
\item[] $I:$  inclination of the orbital plane
\item[] $\Omega:$  longitude of the ascending node
\item[] $\omega:$  argument of pericenter
\item[] $t_p:$ time of pericenter passage
\item[] $t_0:$ reference epoch
\item[] $\mathcal{M}\doteq \nk\ton{t - t_p}:$ mean anomaly
\item[] $\eta\doteq\nk\ton{t_0-t_p}:$ mean anomaly at epoch
\item[] $f:$  true anomaly
\item[] $f_0:$  true anomaly at epoch
\item[] $\Delta U:$ extra-potential of the modified model of gravity
\item[] $\bds A:$ disturbing acceleration
\item[] $A_R:$ radial component of $\bds A$
\item[] $A_T:$ transverse component of $\bds A$
\item[] $A_N:$ normal component of $\bds A$
\end{description}
\section{Computational scheme}\lb{analy}
If the motion of a test particle about its primary is affected by some relatively small post-Keplerian (pK) acceleration $\bds A$ of arbitrary origin, the impact of the latter on the otherwise Keplerian trajectory of the orbiter can be calculated perturbatively as follows.
\citet{1993CeMDA..55..209C}, working in the $RTN$ frame, analytically calculated the instantaneous perturbations $\Delta R\ton{f},~\Delta T\ton{f},~\Delta N\ton{f}$ of the radial, transverse and normal components
$R,~T,~N$  of the position  vector $\bds r$ induced by a generic disturbing acceleration $\bds A$: they are
\begin{align}
\Delta R\ton{f} \lb{rR}&= \rp{r\ton{f}}{a}\Delta a\ton{f} -a\cos f\Delta e\ton{f} +\rp{ae\sin f}{\sqrt{1-e^2}}\Delta\mathcal{M}\ton{f}, \\ \nonumber \\
\Delta T\ton{f} \lb{rT}&= a\sin f\qua{1 + \rp{r\ton{f}}{p}}\Delta e\ton{f} + r\ton{f}\qua{\cI\Delta\Omega\ton{f}+\Delta\omega\ton{f}} +\rp{a^2}{r\ton{f}}\sqrt{1-e^2}\Delta\mathcal{M}\ton{f}, \\ \nonumber \\
\Delta N\ton{f} \lb{rN}&= r\ton{f}\qua{\sin u~\Delta I\ton{f} -\sI\cos u~\Delta\Omega\ton{f}}.
\end{align}
In \rfrs{rR}{rN}, the instantaneous changes $\Delta a\ton{f},~\Delta e\ton{f},~\Delta I\ton{f},~\Delta\Omega\ton{f},~\Delta\omega\ton{f}$  must be worked out as
\eqi
\Delta\kappa\ton{f}=\int_{f_0}^f\dert{\kappa}{t} \dert{t}{f^{'}} df^{'},~\kappa=a,~e,~I,~\Omega,~\omega,\lb{Dk}
\eqf
where the time derivatives $d\kappa/dt$ of the osculating Keplerian orbital elements $\kappa$ are to be taken from the right-hand-sides of the Gauss equations
\begin{align}
\dert a t \lb{dadt}& = \rp{2}{\nk\sqrt{1-e^2}}\qua{e A_R \snf + A_T\ton{\rp{p}{r}}}, \\ \nonumber \\
\dert e t \lb{dedt}& = \rp{\sqrt{1-e^2}}{\nk a}\grf{A_R \snf + A_T\qua{\csf + \rp{1}{e}\ton{1 - \rp{r}{a}} } }, \\ \nonumber \\
\dert I t \lb{dIdt}& = \rp{1}{\nk a \sqrt{1 - e^2}}A_N\ton{\rp{r}{a}}\cu, \\ \nonumber \\
\dert \Omega t \lb{dOdt}& = \rp{1}{\nk a \sI\sqrt{1 - e^2}}A_N\ton{\rp{r}{a}}\su, \\ \nonumber \\
\dert \omega t \lb{dodt}& = -\cI\dert\Omega t + \rp{\sqrt{1-e^2}}{\nk a e}\qua{ -A_R\csf + A_T\ton{1 + \rp{r}{p}}\snf },
\end{align}
evaluated onto the  Keplerian ellipse
\eqi
r=\rp{p}{1+e\cos f}\lb{Kepless}
\eqf
as unperturbed reference trajectory; the same holds also for
\eqi
\dert t f = \rp{r^2}{\sqrt{\mu p}}= \rp{\ton{1-e^2}^{3/2}}{\nk\ton{1+e\cos f}^2}\lb{dtdfKep}
\eqf
entering \rfr{Dk}. The case of the mean anomaly $\mathcal{M}$ is subtler; it requires more care.
Indeed, if the mean motion $\nk$ is time-dependent because of some physical phenomena, it can be written as\footnote{The mean anomaly at epoch is denoted as $\eta$ by \citet{Nobilibook87}, $l_0$ by \citet{1991ercm.book.....B}, and $\epsilon^{'}$ by \citet{2003ASSL..293.....B}. It is a \virg{slow} variable in the sense that its time derivative vanishes in the limit $\bds A\rightarrow 0$; cfr. with \rfr{detadt}. } \citep{Nobilibook87,1991ercm.book.....B,2003ASSL..293.....B}
\eqi
\mathcal{M}\ton{t} = \eta + \int_{t_0}^{t} \nk\ton{t^{'}}dt^{'};\lb{Mt}
\eqf
the Gauss equation for the variation of the mean anomaly at epoch is
\citep{Nobilibook87,1991ercm.book.....B,2003ASSL..293.....B}
\eqi
\dert\eta t = - \rp{2}{\nk a}A_R\ton{\rp{r}{a}} -\rp{\ton{1-e^2}}{\nk a e}\qua{ -A_R\csf + A_T\ton{1 + \rp{r}{p}}\snf }\lb{detadt}.
\eqf
If $\nk$ is constant, as in the Keplerian case, \rfr{Mt} reduces to the usual form
\eqi
\mathcal{M}\ton{t}= \eta + \nk\ton{t-t_0}.
\eqf
In general, when a disturbing acceleration is present, the semimajor axis $a$ varies according to \rfr{dadt}; thus, also the mean motion $\nk$ experiences a change\footnote{We neglect the case $\mu\ton{t}$.}
\eqi
\nk\rightarrow \nk+\Delta\nk\ton{t}
\eqf
which can be calculated in terms of the true anomaly $f$ as
\eqi
\Delta \nk\ton{f}=\derp{\nk}{a}\Delta a\ton{f}= -\rp{3}{2}\rp{\nk}{a}\int_{f_0}^f\dert a t \dert{t}{f^{'}}df^{'}\lb{Dn}
\eqf
by means of \rfr{dadt} and \rfr{dtdfKep}.
Depending on the specific perturbation considered, \rfr{Dn} does not generally vanish.
Thus, the total change experienced by the mean anomaly $\mathcal{M}$ due to the disturbing acceleration $\bds A$ can be obtained as
\eqi
\Delta\mathcal{M}\ton{f} = \Delta\eta\ton{f} + \int_{t_0}^{t}\Delta\nk\ton{t^{'}} dt^{'},\lb{anom}
\eqf
where
\begin{align}
\Delta\eta\ton{f} &= \int_{f_0}^f\dert\eta t \dert{t}{f^{'}} df^{'}, \\ \nonumber \\
\int_{t_0}^{t}\Delta\nk\ton{t^{'}} dt^{'} \lb{inte}& = -\rp{3}{2}\rp{\nk}{a}\int_{f_0}^f\Delta a\ton{f^{'}}\dert{t}{f^{'}}df^{'}.
\end{align}
It should be stressed that, depending on the specific perturbing acceleration $\bds A$ at hand, the calculation of \rfr{inte} may turn out to be rather cumbersome.
\section{Tables and Figures}\lb{tabfig}
\begin{table*}
\caption{Relevant orbital parameters of the existing geodetic satellites of the LAGEOS family.}\lb{tavola0}
\begin{center}
\begin{tabular}{|l|l|l|l|}
  \hline
   & $a$ (km) & $e$ & $\Pb$ (hr)\\
\hline
LAGEOS & $12270$ & $0.0045$  & $3.7573$ \\
LAGEOS II & $12163$ & $0.0135$  & $3.7085$ \\
LARES & $7828$ & $0.0008$ &  $1.9146$ \\
\hline
\end{tabular}
\end{center}
\end{table*}
%
%

%
%

%
%
\begin{table}[htp]
\begin{center}
\begin{tabular}{|c|c|c|} \hline
Model & Parameter & Constraint \\ \hline
 Reissner-Nordstr\"om \citep{wald2010general} &  $\left|Q\right| = 2c\sqrt{\rp{2\uppi \varepsilon_0 K_2}{G}}$ & $\lesssim 7.93\times 10^{11}~\textrm{C}$, from \rfr{ReNo}\\
 \citep{2012MNRAS.427.1555I} $f(T)$ & $\left|\alpha\right|=\rp{\left|K_2\right|}{3c^2}$ & $\lesssim 7.79 \times 10^{-12}~\textrm{m}^2$, from \rfr{bound1}\\
 \citep{2015PhRvD..91j4014R} $f(T)$ & $\left|\alpha\right|=\rp{\left|K_2\right|}{16 c^2}$ & $\lesssim 1.46 \times 10^{-12}~\textrm{m}^2$, from \rfr{bound2}\\
  \citep{Maeda:2006hj} Einstein-Gauss-Bonnet & $\left|\tilde{q}\right|=\rp{c^2\left|K_2\right|}{2\mu^2}$ & $\lesssim 5.94\times 10^{-7}$, from \rfr{K2}\\
  \citep{Ali:2015tva}  & $\left|\eta\right|=\rp{2c\left|K_2\right|}{\hbar G}$ & $\lesssim 1.79\times 10^{59}$, from \rfr{K2}\\
   \citep{Chakraborty:2016lxo} Kalb-Ramond & $\left|b\right| = \rp{3\left|K_3\right|}{\mu}$ & $\lesssim 0.038~\textrm{m}^2$, from \rfr{boundK3}\\  \hline
    \end{tabular}
\end{center}\caption{Constraints on the parameters of the various models treated in Sections~\ref{secU2}~to~\ref{secU3} in terms of the phenomenological ones on $K_2,~K_3$ inferred from the SLR data of the LAGEOS satellites.}
\label{tabletot}
\end{table}%

\bibliography{PXbib,IorioFupeng,fTbib,semimabib,bibML}{}

\begin{thebibliography}{}

\bibitem[\protect\citeauthoryear{{Abbott}, {Abbott}, {Abbott}, {Abernathy},
  {Acernese}, {Ackley}, {Adams}, {Adams}, {Addesso}, {Adhikari} \& et
  al.}{{Abbott} et~al.}{2016}]{GW2016}
{Abbott} B.~P.,  {Abbott} R.,  {Abbott} T.~D.,  {Abernathy} M.~R.,  {Acernese}
  F.,  {Ackley} K.,  {Adams} C.,  {Adams} T.,  {Addesso} P.,  {Adhikari} R.~X.,
     et al. 2016, Physical Review Letters, 116, 061102

\bibitem[\protect\citeauthoryear{{Abbott}, {Abbott}, {Abbott}, {Acernese},
  {Ackley}, {Adams}, {Adams}, {Addesso}, {Adhikari}, {Adya} \& et al.}{{Abbott}
  et~al.}{2017}]{GW2017}
{Abbott} B.~P.,  {Abbott} R.,  {Abbott} T.~D.,  {Acernese} F.,  {Ackley} K.,
  {Adams} C.,  {Adams} T.,  {Addesso} P.,  {Adhikari} R.~X.,  {Adya} V.~B.,
  et al. 2017, Physical Review Letters, 119, 161101

\bibitem[\protect\citeauthoryear{{Adelberger}, {Gundlach}, {Heckel}, {Hoedl} \&
  {Schlamminger}}{{Adelberger} et~al.}{2009}]{2009PrPNP..62..102A}
{Adelberger} E.~G.,  {Gundlach} J.~H.,  {Heckel} B.~R.,  {Hoedl} S.,
  {Schlamminger} S.,  2009, Progr. Part. Nucl. Phys., 62, 102

\bibitem[\protect\citeauthoryear{{Adelberger}, {Heckel}, {Hoedl}, {Hoyle},
  {Kapner} \& {Upadhye}}{{Adelberger} et~al.}{2007}]{2007PhRvL..98m1104A}
{Adelberger} E.~G.,  {Heckel} B.~R.,  {Hoedl} S.,  {Hoyle} C.~D.,  {Kapner}
  D.~J.,    {Upadhye} A.,  2007, Phys. Rev. Lett., 98, 131104

\bibitem[\protect\citeauthoryear{{Adelberger}, {Heckel} \&
  {Nelson}}{{Adelberger} et~al.}{2003}]{2003ARNPS..53...77A}
{Adelberger} E.~G.,  {Heckel} B.~R.,    {Nelson} A.~E.,  2003, Annu. Rev. Nucl.
  Part. S., 53, 77

\bibitem[\protect\citeauthoryear{Aldrovandi \& Pereira}{Aldrovandi \&
  Pereira}{2012}]{pereira}
Aldrovandi R.,  Pereira J.~G.,  2012, Teleparallel gravity: an introduction.
Vol.~173, Springer Science \& Business Media

\bibitem[\protect\citeauthoryear{Ali \& Khalil}{Ali \&
  Khalil}{2016}]{Ali:2015tva}
Ali A.~F.,  Khalil M.~M.,  2016, Nucl. Phys., B909, 173

\bibitem[\protect\citeauthoryear{Allemandi \& Ruggiero}{Allemandi \&
  Ruggiero}{2007}]{Allemandi:2006bm}
Allemandi G.,  Ruggiero M.~L.,  2007, Gen. Rel. Grav., 39, 1381

\bibitem[\protect\citeauthoryear{Antoniou, Bakopoulos \& Kanti}{Antoniou
  et~al.}{2018a}]{Antoniou:2017hxj}
Antoniou G.,  Bakopoulos A.,    Kanti P.,  2018a, Phys. Rev., D97, 084037

\bibitem[\protect\citeauthoryear{Antoniou, Bakopoulos \& Kanti}{Antoniou
  et~al.}{2018b}]{Antoniou:2017acq}
Antoniou G.,  Bakopoulos A.,    Kanti P.,  2018b, Phys. Rev. Lett., 120, 131102

\bibitem[\protect\citeauthoryear{{Appleby}, {Rodr{\'{\i}}guez} \&
  {Altamimi}}{{Appleby} et~al.}{2016}]{2016JGeod..90.1371A}
{Appleby} G.,  {Rodr{\'{\i}}guez} J.,    {Altamimi} Z.,  2016, J. Geod., 90,
  1371

\bibitem[\protect\citeauthoryear{Astier et~al.,}{Astier
  et~al.}{2006}]{Astier:2005qq}
Astier P.,  et~al., 2006, Astron. Astrophys., 447, 31

\bibitem[\protect\citeauthoryear{Barris et~al.,}{Barris
  et~al.}{2004}]{Barris:2003dq}
Barris B.~J.,  et~al., 2004, Astrophys. J., 602, 571

\bibitem[\protect\citeauthoryear{Berry \& Gair}{Berry \&
  Gair}{2011}]{Berry:2011pb}
Berry C. P.~L.,  Gair J.~R.,  2011, Phys. Rev., D83, 104022

\bibitem[\protect\citeauthoryear{Berti et~al.,}{Berti
  et~al.}{2015}]{Berti:2015itd}
Berti E.,  et~al., 2015, Class. Quant. Grav., 32, 243001

\bibitem[\protect\citeauthoryear{{Bertotti}, {Farinella} \&
  {Vokrouhlick\'{y}}}{{Bertotti} et~al.}{2003}]{2003ASSL..293.....B}
{Bertotti} B.,  {Farinella} P.,    {Vokrouhlick\'{y}} D.,  2003, {Physics of
  the Solar System}.
Kluwer, Dordrecht

\bibitem[\protect\citeauthoryear{Bhattacharya \& Chakraborty}{Bhattacharya \&
  Chakraborty}{2017}]{Bhattacharya:2016naa}
Bhattacharya S.,  Chakraborty S.,  2017, Phys. Rev., D95, 044037

\bibitem[\protect\citeauthoryear{Bonanno \& Reuter}{Bonanno \&
  Reuter}{2000}]{Bonanno:2000ep}
Bonanno A.,  Reuter M.,  2000, Phys. Rev., D62, 043008

\bibitem[\protect\citeauthoryear{{Brumberg}}{{Brumberg}}{1991}]{1991ercm.book.....B}
{Brumberg} V.~A.,  1991, {Essential Relativistic Celestial Mechanics}.
Adam Hilger, Bristol

\bibitem[\protect\citeauthoryear{Cai, Capozziello, De~Laurentis \&
  Saridakis}{Cai et~al.}{2016}]{Cai:2015emx}
Cai Y.-F.,  Capozziello S.,  De~Laurentis M.,    Saridakis E.~N.,  2016, Rept.
  Prog. Phys., 79, 106901

\bibitem[\protect\citeauthoryear{Capone \& Ruggiero}{Capone \&
  Ruggiero}{2010}]{Capone:2009xk}
Capone M.,  Ruggiero M.~L.,  2010, Class. Quant. Grav., 27, 125006

\bibitem[\protect\citeauthoryear{Capozziello \& De~Laurentis}{Capozziello \&
  De~Laurentis}{2011}]{Capozziello:2011et}
Capozziello S.,  De~Laurentis M.,  2011, Phys. Rept., 509, 167

\bibitem[\protect\citeauthoryear{Capozziello, Stabile \& Troisi}{Capozziello
  et~al.}{2006}]{Capozziello:2006jj}
Capozziello S.,  Stabile A.,    Troisi A.,  2006, Mod. Phys. Lett., A21, 2291

\bibitem[\protect\citeauthoryear{Capozziello, Stabile \& Troisi}{Capozziello
  et~al.}{2007}]{Capozziello:2007ms}
Capozziello S.,  Stabile A.,    Troisi A.,  2007, Phys. Rev., D76, 104019

\bibitem[\protect\citeauthoryear{{Casotto}}{{Casotto}}{1993}]{1993CeMDA..55..209C}
{Casotto} S.,  1993, Celest. Mech. Dyn. Astr., 55, 209

\bibitem[\protect\citeauthoryear{{Cervantes-Cota}, {Galindo-Uribarri} \&
  {Smoot}}{{Cervantes-Cota} et~al.}{2016}]{2016Univ....2...22C}
{Cervantes-Cota} J.,  {Galindo-Uribarri} S.,    {Smoot} G.,  2016, Universe, 2,
  22

\bibitem[\protect\citeauthoryear{Chakraborty \& SenGupta}{Chakraborty \&
  SenGupta}{2017}]{Chakraborty:2016lxo}
Chakraborty S.,  SenGupta S.,  2017, JCAP, 1707, 045

\bibitem[\protect\citeauthoryear{Clifton, Ferreira, Padilla \& Skordis}{Clifton
  et~al.}{2012}]{Clifton:2011jh}
Clifton T.,  Ferreira P.~G.,  Padilla A.,    Skordis C.,  2012, Phys. Rept.,
  513, 1

\bibitem[\protect\citeauthoryear{{Combrinck}}{{Combrinck}}{2010}]{Combrinck2010}
{Combrinck} L.,  2010, in {{Xu}, G.} ed., {Sciences of Geodesy - I } {Satellite
  Laser Ranging}.
Springer, Berlin, Heidelberg, pp 301--338

\bibitem[\protect\citeauthoryear{{Coulot}, {Berio}, {Laurain}, {F\'{e}raudy},
  {Exertier} \& {Deleflie}}{{Coulot} et~al.}{2008}]{slr2006}
{Coulot} D.,  {Berio} P.,  {Laurain} O.,  {F\'{e}raudy} D.,  {Exertier} P.,
  {Deleflie} F.,  2008, in {Luck} J.,  {Moore} C.,   {Wilson} P.,  eds,
  {Proceedings of the 15th International Workshop on Laser Ranging, Canberra,
  Australia, October 2006} {Analysis of 13 years (1993-2005) of Satellite Laser
  Ranging data on the two LAGEOS satellites for Terrestrial Reference Frames
  and Earth Orientation Parameters}.
EOS Space Systems, Canberra, pp 120--130

\bibitem[\protect\citeauthoryear{De~Felice \& Tsujikawa}{De~Felice \&
  Tsujikawa}{2009}]{DeFelice:2008wz}
De~Felice A.,  Tsujikawa S.,  2009, Phys. Lett., B675, 1

\bibitem[\protect\citeauthoryear{De~Felice \& Tsujikawa}{De~Felice \&
  Tsujikawa}{2010}]{DeFelice:2010aj}
De~Felice A.,  Tsujikawa S.,  2010, Living Rev. Rel., 13, 3

\bibitem[\protect\citeauthoryear{{de Martino}, {De Laurentis} \&
  {Capozziello}}{{de Martino} et~al.}{2015}]{2015Univ....1..123D}
{de Martino} I.,  {De Laurentis} M.,    {Capozziello} S.,  2015, Universe, 1

\bibitem[\protect\citeauthoryear{de Rham}{de~Rham}{2014}]{deRham:2014zqa}
de Rham C.,  2014, Living Rev. Rel., 17, 7

\bibitem[\protect\citeauthoryear{{Debono} \& {Smoot}}{{Debono} \&
  {Smoot}}{2016}]{2016Univ....2...23D}
{Debono} I.,  {Smoot} G.~F.,  2016, Universe, 2, 23

\bibitem[\protect\citeauthoryear{{Dobrescu} \& {Mocioiu}}{{Dobrescu} \&
  {Mocioiu}}{2006}]{2006JHEP...11..005D}
{Dobrescu} B.~A.,  {Mocioiu} I.,  2006, J. High Energy Phys., 11, 005

\bibitem[\protect\citeauthoryear{Eisenstein et~al.,}{Eisenstein
  et~al.}{2005}]{Eisenstein:2005su}
Eisenstein D.~J.,  et~al., 2005, Astrophys. J., 633, 560

\bibitem[\protect\citeauthoryear{{Farrugia}, {Said} \& {Ruggiero}}{{Farrugia}
  et~al.}{2016}]{2016PhRvD..93j4034F}
{Farrugia} G.,  {Said} J.~L.,    {Ruggiero} M.~L.,  2016, Phys. Rev. D, 93,
  104034

\bibitem[\protect\citeauthoryear{Ferraro \& Fiorini}{Ferraro \&
  Fiorini}{2008}]{Ferraro:2008ey}
Ferraro R.,  Fiorini F.,  2008, Phys. Rev., D78, 124019

\bibitem[\protect\citeauthoryear{{Ferrer} \& {Grifols}}{{Ferrer} \&
  {Grifols}}{1998}]{1998PhRvD..58i6006F}
{Ferrer} F.,  {Grifols} J.~A.,  1998, Phys. Rev. D, 58, 096006

\bibitem[\protect\citeauthoryear{{Ferrer} \& {Nowakowski}}{{Ferrer} \&
  {Nowakowski}}{1999}]{1999PhRvD..59g5009F}
{Ferrer} F.,  {Nowakowski} M.,  1999, Phys. Rev. D, 59, 075009

\bibitem[\protect\citeauthoryear{{Fischbach}, {Krause}, {Mostepanenko} \&
  {Novello}}{{Fischbach} et~al.}{2001}]{2001PhRvD..64g5010F}
{Fischbach} E.,  {Krause} D.~E.,  {Mostepanenko} V.~M.,    {Novello} M.,  2001,
  Phys. Rev. D, 64, 075010

\bibitem[\protect\citeauthoryear{Hinshaw et~al.,}{Hinshaw
  et~al.}{2013}]{Hinshaw:2012aka}
Hinshaw G.,  et~al., 2013, Astrophys. J. Suppl., 208, 19

\bibitem[\protect\citeauthoryear{Horndeski}{Horndeski}{1974}]{Horndeski:1974wa}
Horndeski G.~W.,  1974, Int. J. Theor. Phys., 10, 363

\bibitem[\protect\citeauthoryear{Iorio}{Iorio}{2012}]{Iorio:2011aa}
Iorio L.,  2012, Gen. Rel. Grav., 44, 1753

\bibitem[\protect\citeauthoryear{{Iorio}}{{Iorio}}{2012}]{2012AnP...524..371I}
{Iorio} L.,  2012, Ann. Phys.-Berlin, 524, 371

\bibitem[\protect\citeauthoryear{{Iorio}, {Radicella} \& {Ruggiero}}{{Iorio}
  et~al.}{2015}]{2015JCAP...08..021I}
{Iorio} L.,  {Radicella} N.,    {Ruggiero} M.~L.,  2015, J. Cosmol. Astropart.
  Phys., 8, 021

\bibitem[\protect\citeauthoryear{Iorio, Ruggiero, Radicella \& Saridakis}{Iorio
  et~al.}{2016}]{Iorio:2016sqy}
Iorio L.,  Ruggiero M.~L.,  Radicella N.,    Saridakis E.~N.,  2016, Phys. Dark
  Univ., 13, 111

\bibitem[\protect\citeauthoryear{{Iorio} \& {Saridakis}}{{Iorio} \&
  {Saridakis}}{2012}]{2012MNRAS.427.1555I}
{Iorio} L.,  {Saridakis} E.~N.,  2012, MNRAS, 427, 1555

\bibitem[\protect\citeauthoryear{Jusufi}{Jusufi}{2017}]{Jusufi:2016sym}
Jusufi K.,  2017, Int. J. Geom. Meth. Mod. Phys., 14, 1750137

\bibitem[\protect\citeauthoryear{Kanti, Kleihaus \& Kunz}{Kanti
  et~al.}{2011}]{Kanti:2011jz}
Kanti P.,  Kleihaus B.,    Kunz J.,  2011, Phys. Rev. Lett., 107, 271101

\bibitem[\protect\citeauthoryear{Kanti, Kleihaus \& Kunz}{Kanti
  et~al.}{2012}]{Kanti:2011yv}
Kanti P.,  Kleihaus B.,    Kunz J.,  2012, Phys. Rev., D85, 044007

\bibitem[\protect\citeauthoryear{Knop et~al.,}{Knop
  et~al.}{2003}]{Knop:2003iy}
Knop R.~A.,  et~al., 2003, Astrophys. J., 598, 102

\bibitem[\protect\citeauthoryear{{Kramer}, {Stairs}, {Manchester},
  {McLaughlin}, {Lyne}, {Ferdman}, {Burgay}, {Lorimer}, {Possenti}, {D'Amico},
  {Sarkissian}, {Hobbs}, {Reynolds}, {Freire} \& {Camilo}}{{Kramer}
  et~al.}{2006}]{pulsar2006}
{Kramer} M.,  {Stairs} I.~H.,  {Manchester} R.~N.,  {McLaughlin} M.~A.,  {Lyne}
  A.~G.,  {Ferdman} R.~D.,  {Burgay} M.,  {Lorimer} D.~R.,  {Possenti} A.,
  {D'Amico} N.,  {Sarkissian} J.~M.,  {Hobbs} G.~B.,  {Reynolds} J.~E.,
  {Freire} P.~C.~C.,    {Camilo} F.,  2006, Science, 314, 97

\bibitem[\protect\citeauthoryear{{Lake}}{{Lake}}{2016}]{2016Univ....2...24L}
{Lake} M.,  2016, Universe, 2, 24

\bibitem[\protect\citeauthoryear{Lin, Zhai \& Li}{Lin
  et~al.}{2017}]{Lin:2016nvj}
Lin R.-H.,  Zhai X.-H.,    Li X.-Z.,  2017, Eur. Phys. J., C77, 504

\bibitem[\protect\citeauthoryear{Linder}{Linder}{2010}]{Linder:2010py}
Linder E.~V.,  2010, Phys. Rev., D81, 127301

\bibitem[\protect\citeauthoryear{Maeda \& Dadhich}{Maeda \&
  Dadhich}{2007}]{Maeda:2006hj}
Maeda H.,  Dadhich N.,  2007, Phys. Rev., D75, 044007

\bibitem[\protect\citeauthoryear{{Maluf}}{{Maluf}}{2016}]{2016Univ....2...19M}
{Maluf} J.,  2016, Universe, 2, 19

\bibitem[\protect\citeauthoryear{Maluf}{Maluf}{2013}]{Maluf:2013gaa}
Maluf J.~W.,  2013, Annalen Phys., 525, 339

\bibitem[\protect\citeauthoryear{Martin}{Martin}{2012}]{Martin:2012bt}
Martin J.,  2012, Comptes Rendus Physique, 13, 566

\bibitem[\protect\citeauthoryear{{Milani}, {Nobili} \& {Farinella}}{{Milani}
  et~al.}{1987}]{Nobilibook87}
{Milani} A.,  {Nobili} A.,    {Farinella} P.,  1987, {Non-gravitational
  perturbations and satellite geodesy}.
Adam Hilger, Bristol

\bibitem[\protect\citeauthoryear{{Mostepanenko} \& {Sokolov}}{{Mostepanenko} \&
  {Sokolov}}{1987}]{1987PhLA..125..405M}
{Mostepanenko} V.~M.,  {Sokolov} I.~Y.,  1987, Phys. Lett. A, 125, 405

\bibitem[\protect\citeauthoryear{Naruko, Yoshida \& Mukohyama}{Naruko
  et~al.}{2016}]{Naruko:2015zze}
Naruko A.,  Yoshida D.,    Mukohyama S.,  2016, Class. Quant. Grav., 33, 09LT01

\bibitem[\protect\citeauthoryear{Nojiri \& Odintsov}{Nojiri \&
  Odintsov}{2005}]{Nojiri:2005jg}
Nojiri S.,  Odintsov S.~D.,  2005, Phys. Lett., B631, 1

\bibitem[\protect\citeauthoryear{Nunes, Pan \& Saridakis}{Nunes
  et~al.}{2016}]{Nunes:2016qyp}
Nunes R.~C.,  Pan S.,    Saridakis E.~N.,  2016, JCAP, 1608, 011

\bibitem[\protect\citeauthoryear{Peebles \& Ratra}{Peebles \&
  Ratra}{2003}]{Peebles:2002gy}
Peebles P. J.~E.,  Ratra B.,  2003, Rev. Mod. Phys., 75, 559

\bibitem[\protect\citeauthoryear{Perlmutter et~al.,}{Perlmutter
  et~al.}{1998}]{Perlmutter:1997zf}
Perlmutter S.,  et~al., 1998, Nature, 391, 51

\bibitem[\protect\citeauthoryear{{Randall} \& {Sundrum}}{{Randall} \&
  {Sundrum}}{1999}]{1999PhRvL..83.4690R}
{Randall} L.,  {Sundrum} R.,  1999, Phys. Rev. Lett., 83, 4690

\bibitem[\protect\citeauthoryear{Riess et~al.,}{Riess
  et~al.}{1998}]{Riess:1998cb}
Riess A.~G.,  et~al., 1998, Astron. J., 116, 1009

\bibitem[\protect\citeauthoryear{Riess et~al.,}{Riess
  et~al.}{2004}]{Riess:2004nr}
Riess A.~G.,  et~al., 2004, Astrophys. J., 607, 665

\bibitem[\protect\citeauthoryear{Ruggiero \& Iorio}{Ruggiero \&
  Iorio}{2007}]{Ruggiero:2006qv}
Ruggiero M.~L.,  Iorio L.,  2007, JCAP, 0701, 010

\bibitem[\protect\citeauthoryear{{Ruggiero} \& {Radicella}}{{Ruggiero} \&
  {Radicella}}{2015}]{2015PhRvD..91j4014R}
{Ruggiero} M.~L.,  {Radicella} N.,  2015, Phys. Rev. D, 91, 104014

\bibitem[\protect\citeauthoryear{Saridakis \& Tsoukalas}{Saridakis \&
  Tsoukalas}{2016}]{Saridakis:2016ahq}
Saridakis E.~N.,  Tsoukalas M.,  2016, Phys. Rev., D93, 124032

\bibitem[\protect\citeauthoryear{Sotiriou \& Faraoni}{Sotiriou \&
  Faraoni}{2010}]{Sotiriou:2008rp}
Sotiriou T.~P.,  Faraoni V.,  2010, Rev. Mod. Phys., 82, 451

\bibitem[\protect\citeauthoryear{Sotiriou \& Zhou}{Sotiriou \&
  Zhou}{2014}]{Sotiriou:2014pfa}
Sotiriou T.~P.,  Zhou S.-Y.,  2014, Phys. Rev., D90, 124063

\bibitem[\protect\citeauthoryear{Spergel et~al.,}{Spergel
  et~al.}{2007}]{Spergel:2006hy}
Spergel D.~N.,  et~al., 2007, Astrophys. J. Suppl., 170, 377

\bibitem[\protect\citeauthoryear{Stelle}{Stelle}{1977}]{Stelle:1976gc}
Stelle K.~S.,  1977, Phys. Rev., D16, 953

\bibitem[\protect\citeauthoryear{Tapley, Bettadpur, Watkins \& Reigber}{Tapley
  et~al.}{2004}]{tapley2004gravity}
Tapley B.~D.,  Bettadpur S.,  Watkins M.,    Reigber C.,  2004, Geophysical
  Research Letters, 31

\bibitem[\protect\citeauthoryear{{Taylor}, {Fowler} \& {McCulloch}}{{Taylor}
  et~al.}{1979}]{1979Natur.277..437T}
{Taylor} J.~H.,  {Fowler} L.~A.,    {McCulloch} P.~M.,  1979, \nat, 277, 437

\bibitem[\protect\citeauthoryear{Tonry et~al.,}{Tonry
  et~al.}{2003}]{Tonry:2003zg}
Tonry J.~L.,  et~al., 2003, Astrophys. J., 594, 1

\bibitem[\protect\citeauthoryear{{Vishwakarma}}{{Vishwakarma}}{2016}]{2016Univ....2...11V}
{Vishwakarma} R.,  2016, Universe, 2, 11

\bibitem[\protect\citeauthoryear{Wald}{Wald}{2010}]{wald2010general}
Wald R.,  2010, General Relativity.
University of Chicago Press

\bibitem[\protect\citeauthoryear{{Will}}{{Will}}{2015}]{2014arXiv1409.7871W}
{Will} C.~M.,  2015, in {Ashtekar} A.,  {Berger} B.~K.,  {Isenberg} J.,
  {MacCallum} M.,  eds, {General Relativity and Gravitation. A Centennial
  Perspective} {Was Einstein Right? A Centenary Assessment}.
Cambridge University Press, Cambridge, pp 49--96

\bibitem[\protect\citeauthoryear{{Xie} \& {Deng}}{{Xie} \&
  {Deng}}{2013}]{2013MNRAS.433.3584X}
{Xie} Y.,  {Deng} X.-M.,  2013, MNRAS, 433, 3584

\bibitem[\protect\citeauthoryear{{Xu}, {Yu} \& {Wu}}{{Xu}
  et~al.}{2018}]{2018XU}
{Xu} B.,  {Yu} H.,    {Wu} P.,  2018, \apj, 855, 89

\end{thebibliography}


\end{document}